\documentclass[reprint, superscriptaddress, secnumarabic, amssymb, nobibnotes, aps, prl]{revtex4-2}

\usepackage{graphicx}
\usepackage{epstopdf}
\usepackage[T1]{fontenc}
\usepackage[latin9]{inputenc}
\usepackage{amsbsy}
\usepackage{gensymb}
\usepackage{amsmath}
\usepackage{amssymb}
\usepackage{bbm}
\usepackage{braket}
\usepackage{xcolor}
\usepackage{multirow}
\allowdisplaybreaks
\usepackage{graphicx}
\usepackage[colorlinks=true]{hyperref}  
\hypersetup{
    bookmarks=true,         % show bookmarks bar?
    unicode=false,          % non-Latin characters 
    pdftoolbar=true,        % show Acrobat
    pdfmenubar=true,        % show Acrobat 
    pdffitwindow=false,     % window fit to page when opened
    pdfstartview={FitH},    % fits the width of the page to the window
    pdftitle={TaXSi},    % title
    pdfauthor={},     % author
    pdfsubject={},   % subject of the document
    pdfcreator={},   % creator of the document
    pdfproducer={}, % producer of the document
    pdfkeywords={} {} {}, % list of keywords
    pdfnewwindow=true,      % links in new window
    colorlinks=true,       % false: boxed links; true: colored links
    linkcolor=blue, %red,          % color of internal links (change box color with linkbordercolor)
    citecolor=blue,        % color of links to bibliography
    filecolor=magenta,      % color of file links
    urlcolor=blue           % color of external links
} 
\usepackage[normalem]{ulem}

\renewcommand{\approx}{\simeq}

\renewcommand{\vec}[1]{\boldsymbol{#1}}

\begin{document}
	\title{\textrm{Evidence for non-unitary triplet-pairing superconductivity in noncentrosymmetric TaRuSi and comparison with isostructural TaReSi  }}
\author{S.~ Sharma}
\affiliation{Department of Physics and Astronomy, McMaster University, Hamilton, Ontario L8S 4M1, Canada}
\author{Sajilesh K. P.}
\affiliation{Department of Physics, Indian Institute of Science Education and Research Bhopal, Bhopal, 462066, India}
\author{A. D. S. Richards}
\affiliation{Department of Physics and Astronomy, McMaster University, Hamilton, Ontario L8S 4M1, Canada}
\author{J. Gautreau}
\affiliation{Department of Physics and Astronomy, McMaster University, Hamilton, Ontario L8S 4M1, Canada}
\author{M.~Pula}
\affiliation{Department of Physics and Astronomy, McMaster University, Hamilton, Ontario L8S 4M1, Canada}
\author{J.~Beare}
\affiliation{Department of Physics and Astronomy, McMaster University, Hamilton, Ontario L8S 4M1, Canada}
\author{K. M. Kojima}
\affiliation{TRIUMF, Vancouver, British Columbia V6T 2A3, Canada}
\author{S. Yoon}
\affiliation{TRIUMF, Vancouver, British Columbia V6T 2A3, Canada}
\affiliation{Department of Physics, Sungkyunkwan University, Suwon 16419, Korea}
\author{Y. Cai}
\affiliation{TRIUMF, Vancouver, British Columbia V6T 2A3, Canada}
\affiliation{Quantum Matter Institute, The University of British Columbia, Vancouver, BC V6T 1Z4, Canada}
\author{R. K. Kushwaha}
\affiliation{Department of Physics, Indian Institute of Science Education and Research Bhopal, Bhopal, 462066, India}
\author{T. Agarwal}
\affiliation{Department of Physics, Indian Institute of Science Education and Research Bhopal, Bhopal, 462066, India}
\author{E.~S. S\o rensen}
\affiliation{Department of Physics and Astronomy, McMaster University, Hamilton, Ontario L8S 4M1, Canada}
\author{R.~P.~Singh}
\email[]{rpsingh@iiserb.ac.in} 
\affiliation{Department of Physics, Indian Institute of Science Education and Research Bhopal, Bhopal, 462066, India}
\author{G.~M.~Luke}
\email[]{luke@mcmaster.ca}
\affiliation{Department of Physics and Astronomy, McMaster University, Hamilton, Ontario L8S 4M1, Canada}
\affiliation{TRIUMF, Vancouver, British Columbia V6T 2A3, Canada}	
	\date{\today}
	\begin{abstract}
		\begin{flushleft}
		\end{flushleft}
We have studied the superconducting properties of the isostructural ternary noncentrosymmetric superconductors TaXSi (X = Re, Ru) with the help of muon spin rotation/relaxation ($\mu$SR) and density functional theory calculations. Our transverse-field $\mu$SR measurements indicate isotropic s-wave superconductivity in TaReSi and multi-gap superconductivity in TaRuSi.  Zero-field $\mu$SR measurements, highly sensitive to very small magnetic fields, find no evidence for spontaneous fields in the superconducting state of TaReSi, whereas we observe small spontaneous fields that onset with superconductivity indicating broken time-reversal symmetry (TRS) superconductivity in TaRuSi.  Using density functional theory calculations, we find that spin-orbit coupling is relatively weak in TaRuSi and strong in TaReSi. Using symmetry analysis, we attribute the broken time-reversal symmetry (TRS) in TaRuSi  to a non-unitary triplet pairing state. Such a state is not allowed in the presence of strong spin-orbit coupling: our finding of no evidence for broken TRS in TaReSi is consistent with this expectation.
%while in TaReSi, this state is suppressed due to strong anti-symmetric spin-orbit coupling.
		
	\end{abstract}
	\maketitle
	
Understanding the origins of unconventional superconductivity has remained a key challenge, despite decades-long research efforts. The role of the underlying crystal structure has been recognized for its role in determining the allowed symmetries of the superconducting state \cite{Sigrist_Ueda_1991}. The discovery of coexisting heavy fermion superconductivity and antiferromagnetism in CePt$ _{3} $Si \cite{Bauer2004,CPS}, showed the potential importance of the absence of inversion symmetry for selecting the superconducting and magnetic state.   In superconductors with crystal structures lacking inversion symmetry: i.e., noncentrosymmetric superconductors (NCS), parity is not a good quantum number, and electronic antisymmetric spin-orbit coupling (ASOC) is allowed by symmetry. This relaxed symmetry in NCS systems, in the presence of strong ASOC, can give rise to the formation of a mixture of spin-singlet and spin-triplet Cooper pairs \cite{rashba1,rashba2,rashba3,EBA,vm}.  Such a superconducting state can go beyond conventional BCS-like superconductivity with various exotic features like time-reversal symmetry (TRS) breaking, zeros or multiple gaps in the energy spectrum, and topologically protected nontrivial surface states \cite{LI,kv,fujimoto,topo1,topo3,Hsoc1,Hsoc3,topo2}. Nevertheless, there are multiple examples of unconventional superconductors with weak ASOC that break time-reversal symmetry, for example, LaNiC$ _{2} $ \cite{LNC1}, La$ _{7} $X$  _{3} $ ($X$ = Ir, Rh, Ni) \cite{LI,LaRh,LN}, SrPtAs \cite{SPA}, CaPtAs \cite{CPA}. In particular, the point-group symmetry of LaNiC$_2$, which is also the point group of TaXSi, dictates that the only superconducting states consistent with time-reversal symmetry are non-unitary triplet pairing states \cite{LNC1}. It was argued from symmetry analysis that TRS breaking in the superconducting state of these systems is only possible in the limit of weak spin-orbit coupling~\cite{Quintanilla_Cywinski_2010}.

Moreover, the nature of superconductivity, affected by tuning the strength of ASOC, where the crystal structure remain the same, has not been widely studied thus far.  TaXSi  (X =Ru/Re) is an ideal candidate system for such a study due to the presence of heavier Re atoms and lighter Ru atoms which leads to varying levels of spin-orbit coupling. 
The presence of heavier Re atoms in the crystal structure of TaReSi, which contributes to the higher ASOC, highlights other aspects of the discussion on unconventional superconductivity, with higher Re concentration being considered as the driving factor for the unconventional nature of some superconductors \cite{Re}. An interesting case study  is available on the $ \alpha $-Mn structured Re$ _{6} $X family where the TRS broken state is observed irrespective of the X element \cite{RZ3,RH2,RT} while a few other compounds from the same family with lesser Re concentration have shown preserved TRS \cite{RW,RTa}. A recent study on the type I superconductor phase of elemental Re attributes the temperature-dependent muon relaxation behavior to muon diffusion  \cite{ReI}, indicating that the nature of the superconducting state (type-I/type-II, TRS) may be highly sensitive to precise sample details in this elemental superconductor.

There are many examples of superconductors with noncentrosymmetric structures that exhibit conventional behavior, raising questions about the selective appearance of non-BCS characteristics \cite{LPS,LPS2,LPG,LRS,LMP,CIS,LS2,LS3,ZI}. Therefore, the current understanding of the NCS superconductors can be improved via further experimental investigations into 
% the role of ASOC in%
 the unconventional nature of superconductors. Here we report the investigations of  TaXSi (X = Re, Ru) via muon spin rotation and relaxation ($\mu$SR) \cite{blundell2021muon}, which is a highly sensitive technique that can detect spontaneous magnetic fields characteristic of broken time-reversal symmetry and the presence/absence of nodes in the superconducting gap. We employ density functional theory to determine the electronic states and the effects of spin-orbit coupling in TaReSi and TaRuSi, which crystallize in the noncentrosymmetric orthorhombic TiFeSi-type structure (space group $Ima2$) \cite{GV,TXS}. This structure falls under the globally stable nonsymmorphic symmetry, which can host nontrivial topological features \cite{nonsym,Leonhardt2021}.

	 The samples for this study were synthesized by arc-melting pure Ta, Re/Ru, and Si powders in stoichiometric ratios under an inert atmosphere. We confirmed bulk superconductivity in our samples with the help of electrical transport, magnetization, and specific heat measurements, which exhibited superconducting transitions at 5.32 K and 3.91 K, respectively, for TaReSi and TaRuSi. More details on the synthesis and characterization of the polycrystalline samples used in this study are described in \cite{TXS}.
	 
	 Muon spectroscopy measurements were performed at M15 and M20 beamlines at TRIUMF's Centre for Molecular and Material Science at Vancouver, which are equipped with a dilution refrigerator and He$^4$ cryostat, respectively.  The samples were cut into small plates with the help of a diamond saw and were mounted on the silver sample holder (cold-finger), utilizing copper grease for good thermal conductivity in the dilution refrigerator and in a helium exchange gas-cooled low background insert in the He$^4$ cryostat. We further secured the samples with a thin silver foil before mounting them into the dilution refrigerator.  We  zeroed the magnetic  field at the sample within an accuracy of 10-20 mG  for zero-field (ZF) $\mu$SR using muonium atoms through the method described by Morris and Heffner \cite{Morris}. The ZF-$\mu$SR measurements were performed in non-spin rotated mode while transverse-field (TF) $\mu$SR measurements were performed in spin-rotated mode, which involves rotating the muon spins perpendicular to the beam velocity before landing in the sample, and the field along the beam direction is applied to the sample. The $\mu$SR data was analyzed with musrfit software \cite{Suter} to obtain physical parameters. 
	
\begin{figure*}
	\centering
	\includegraphics[width=0.82\textwidth]{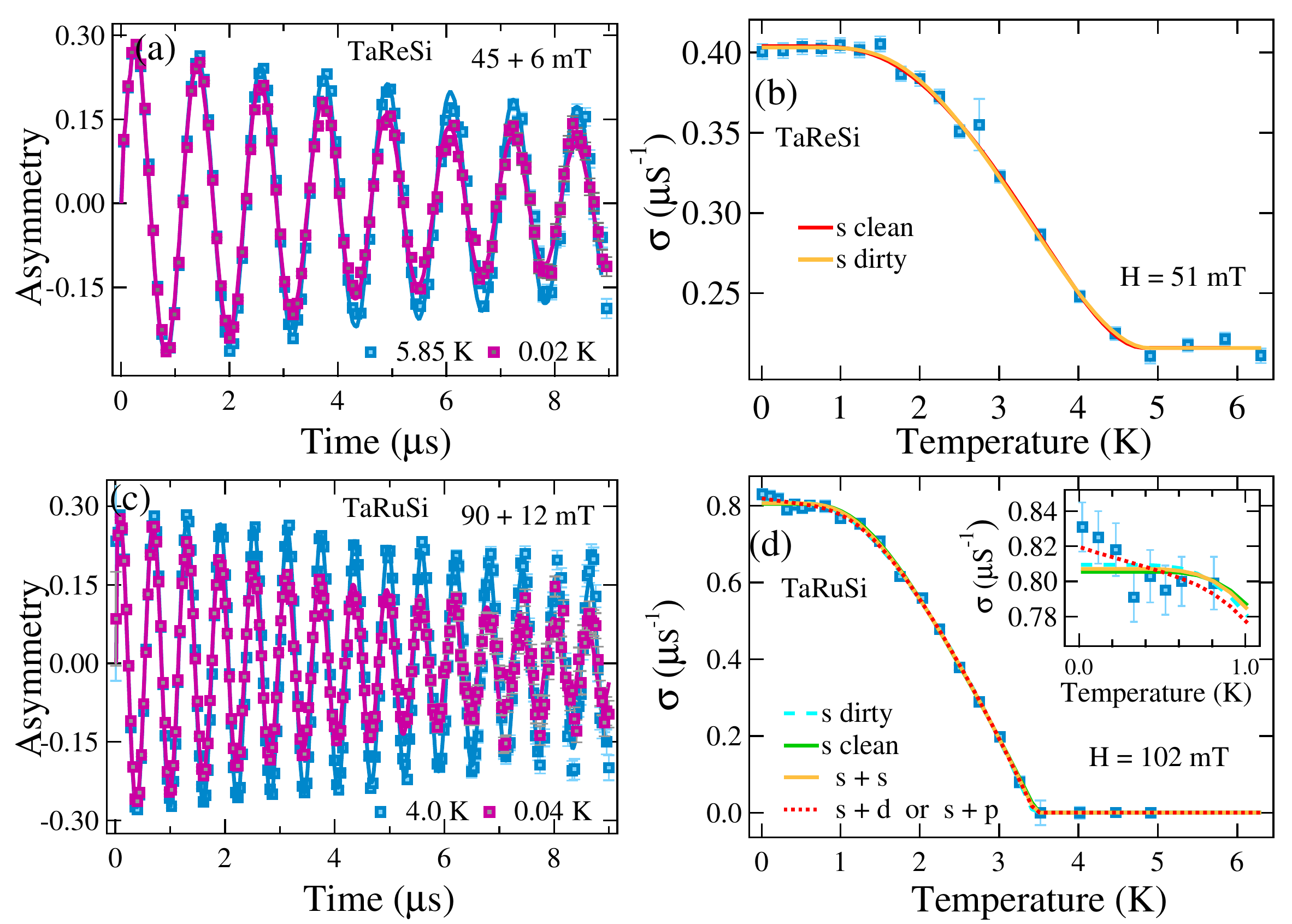}
	\caption{\label{fig1} (a) Transverse field asymmetry spectra in the normal and the superconducting states of TaReSi in rotating reference frame with frequency $\gamma_\mu$45 mT show increased relaxation in the superconducting state due to the formation of flux like lattice. These spectra were measured in the 51 mT transverse field. (b) The temperature dependence of muon depolarisation rate and fit to dirty and clean limit BCS models. (c) Transverse field asymmetry spectra in the normal and superconducting state of TaRuSi in a rotating reference frame with frequency $\gamma_\mu$90 mT. These spectra were measured in the applied transverse field of 102 mT. (d) Fits of the temperature dependence of muon depolarization rate ($\sigma$) at applied fields of 102 mT. The inset shows the relaxation rate curve expanded in the 0 K to 1 K temperature range, demonstrating how the s+d or s+p wave models fit better than the other models. 
  }
\end{figure*}

	Transverse field $ \mu $SR can be used to examine a superconductor in its vortex state to determine the temperature dependence of the magnetic penetration depth and, from that, the energy gap. During the transverse field measurements, we cooled the samples in the 51 mT and 102 mT fields, respectively, for TaReSi and TaRuSi, which were well above the lower critical field for each compound. The field cooling procedure ensures a well-ordered flux line lattice. The application of the magnetic field has reduced the superconducting transition temperature compared to those obtained through zero-field transport measurements. Typical time evolutions of the asymmetry spectra for both compounds are shown in Fig. \ref{fig1}(a and c). For both compounds, the asymmetry spectra are shown in a rotating reference frame for clarity.  
	
	In the mixed state, the presence of a flux line lattice (FLL) creates an inhomogeneous field distribution, which gives rise to a decay of the precession signal as a function of time.  The asymmetry spectra were fit using a two-term sinusoidal decaying function. 
	
\begin{equation}
\begin{split}
G_{\mathrm{TF}}(t) = A[F \;\mathrm{exp}\left(\frac{-\sigma^{2}t^{2}}{2}\right)\mathrm{cos}(\omega_{1}t+\phi)\\+\;(1-F)\;\mathrm{exp}\left(-\psi t\right)\mathrm{cos}(\omega_{2}t+\phi)]
\label{eqn3}
\end{split}
\end{equation}
Here, the first term accounts for the signal from the sample, and the second term accounts for the signal from the silver sample holder (present in the dilution refrigerator data). $F$ is the sample fraction of the signal, while $ \omega_{1} $ and $ \omega_{2} $ correspond to the muon precession frequencies in the sample and the background, respectively, while $A$ is the total asymmetry and $\phi$ is the initial phase of the muons. The depolarization rates for the sample and background signals are denoted as $ \sigma $ and $\psi$. The total sample signal depolarization $ \sigma $ contains the contribution from flux line lattice ($ \sigma_{sc} $) and the small, temperature-independent contribution from randomly oriented nuclear dipole moments ($ \sigma_{N} $), which are added in quadrature. Hence, the contribution from FLL can be obtained as  $\sigma_{sc}$ = $\sqrt{\sigma^{2}-\sigma_{\mathrm{N}}^{2}}$. The superconducting relaxation rate ($\sigma_{sc}$) represents the mean square inhomogeneity in the field, $\langle(\Delta B^2\rangle$, experienced by the muons due to the underlying flux line lattice \cite{Aeppli1987}, where, $\langle(\Delta B)^2\rangle = \langle(B-\langle B\rangle)^2\rangle$, giving the relaxation rate for the FLL
 \begin{equation}
 \label{variance}
     \sigma_{sc}^2 = \gamma_\mu^2\langle(\Delta B)^2\rangle  ,
 \end{equation}
 where $\gamma_\mu$ (= 2$\pi\times$135.5~MHz/T) is muon gyromagnetic ratio. The transverse field relaxation rate $\sigma$ is plotted in Fig.\ref{fig1} (b) and (d) for TaReSi and TaRuSi, respectively. 
 
For small applied fields [$H / H_{c2}  \ll 1$ ], the penetration depth can be calculated from the relaxation rate using Brandt's formulae \cite{brandt} for a triangular Abrikosov vortex lattice: 
	\begin{equation}
    \sigma_{\mathrm{sc}}(T) = \frac{0.0609 \times \gamma_{\mu}\phi_{0}}{\lambda^{2}(T)}.
    \label{braneq2}
    \end{equation}
 Here, $\sigma_{\mathrm{sc}}(T)$ is in $\mu$s$^{-1}$, $\lambda(T)$ is in nm, and $ \phi_{0} $ (2.067$\times$10$^{-15}$ Wb) is the magnetic flux quantum. Hence, the temperature dependence of the relaxation rate is related to that of penetration depth:
\begin{equation}
 \frac{\sigma_{sc}(T)}{\sigma_{sc}(0)} = \frac{\lambda^{-2}(T)}{\lambda^{-2}(0)} ,
\end{equation}
  Within BCS theory the temperature dependence of the energy gap \cite{gapequation} $\Delta{(T,\hat{k})}$, is given by 
\begin{equation}
\Delta(T, \hat{k}) = \Delta(0)\tanh\{1.82\left(1.018\left(\frac{T_c}{T}-1\right)\right)^{0.51}\}g_{\hat{k}}
\label{gap}
\end{equation}
where $\Delta(0)$ is the gap magnitude at zero temperature. 
The term $g_{\hat{k}}$ in equation \ref{gap} accounts for the orientation ($\hat{k}$) dependence of the gap function and can be substituted with 1 for an s-wave model, $\vert\cos(2\phi)\vert$ for a d-wave model, and $\vert\sin(\phi)\vert$ for a p-wave model where $ \phi $ is the azimuthal angle. 
    
	The temperature dependence of the superconducting gap can be obtained from that of penetration depth in the dirty limit using the relation,
	\begin{equation}
 \frac{\lambda^{-2}(T,\hat{k})}{\lambda^{-2}(0)}  = \Biggl\langle\frac{\Delta(T,\hat{k})}{\Delta(0)}\mathrm{tanh}\left[\frac{\Delta(T,\hat{k})}{2k_{B}T}\right]\Biggr\rangle ,
\label{dirty s}
\end{equation} 
and in the clean limit, 
	\begin{equation}
     \frac{\lambda^{-2}(T)}{\lambda^{-2}(0)} = 1+2\Biggl\langle\int_{|\Delta(T,\hat{k})|}^{\infty}\left(\frac{\delta f}{\delta E}\right)\frac{E dE}{\sqrt{E^{2}-\Delta^{2}(T,\hat{k})}}\Biggr\rangle  ,
    \label{eqnclean}
    \end{equation}
where $f = [1+\exp(E/k_{B}T)]^{-1}$ is the Fermi function, and the quantities in the angular brackets are averaged over the Fermi surface.  In order to check for a possible multi-gap nature as was inferred from susceptibility measurements \cite{TXS},  we have also performed a two-gap model fitting, where the total depolarization is expressed as a sum of two components,

\begin{equation}
\frac{\sigma_{FLL}^{-2}(T)}{\sigma_{FLL}^{-2}(0)} = x \frac{\sigma_{FLL}^{-2}(T, \Delta_{0,1})}{\sigma_{FLL}^{-2}(0,\Delta_{0,1})} + (1 - x) \frac{\sigma_{FLL}^{-2}(T, \Delta_{0,2})}{\sigma_{FLL}^{-2}(0,\Delta_{0,2})}
\label{clean s}
\end{equation}
where $ \Delta_{0,1} $ and $ \Delta_{0,2} $ are the gap values at zero temperature and $x$ and $(1-x)$ are the weights corresponding to the two gaps. We have fit the data using s + s, s + p, and s + d wave models. The results of the fit are presented in Table \ref{fit parameters} with the plots shown in Fig. \ref{fig1}(b) and (d) for TaReSi and TaRuSi, respectively. 
 
 \begin{figure*}
 \begin{center}
	\includegraphics[width=\textwidth]{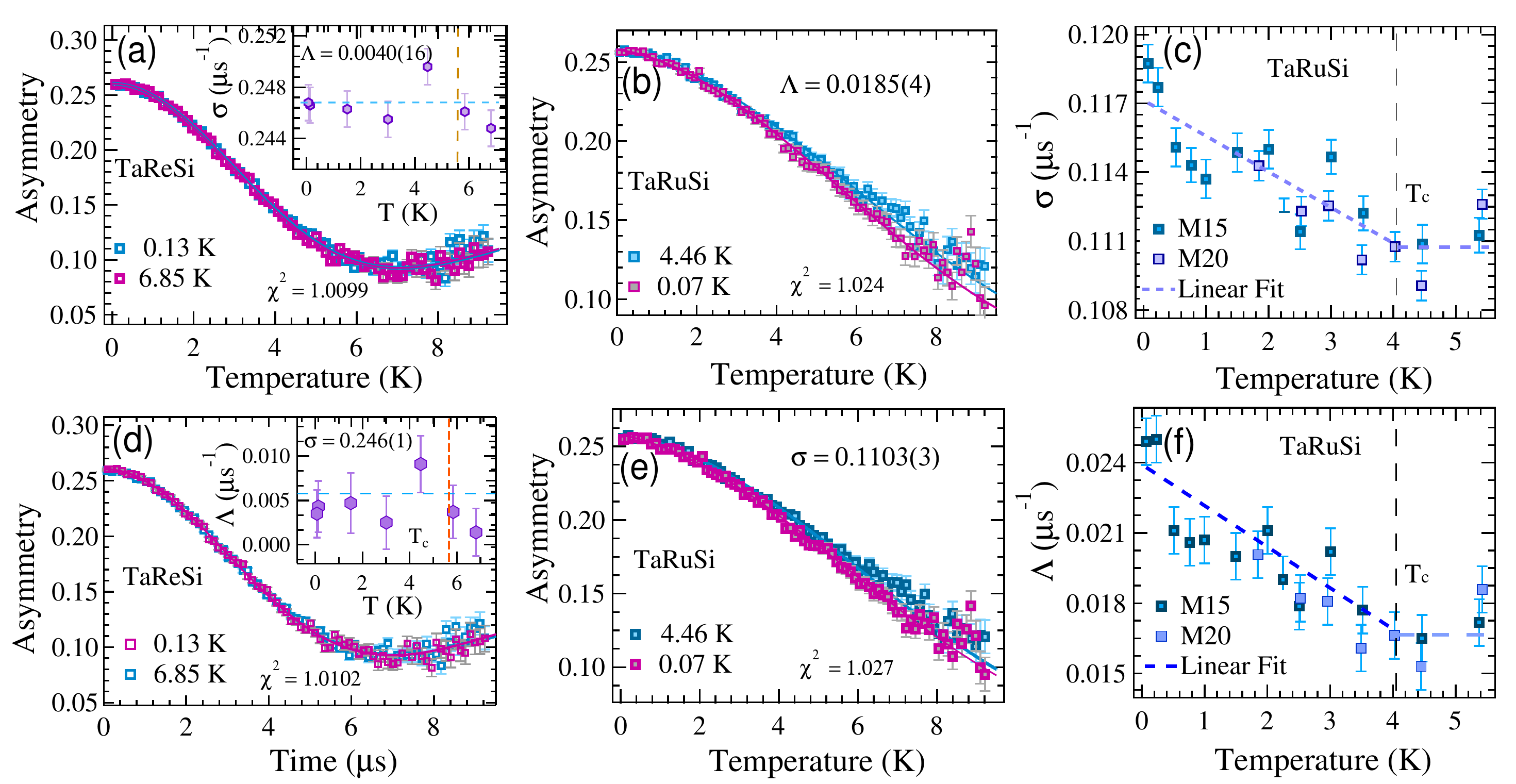}
	\caption{\label{fig2}Zero field muon spectra collected at temperatures above and below T$_{c}$ for TaReSi where asymmetry spectra was fitted with temperature dependent (a) $\sigma$ and (d) $\Lambda$. The spectra for TaReSi have shown no noticeable difference across T$_{c}$ as evident from the relaxation rate([Inset] (a) $\sigma$ and [Inset](d) $\Lambda$ ) versus temperature graph. Zero field muon spectra collected at temperatures above and below T$_{c}$ for TaRuSi is shown where the asymmetry spectra was fitted with temperature dependent (b) $\sigma$ and (e) $\Lambda$ . The temperature dependence of relaxation rate(c) $ \Lambda $  and (f) $\sigma$ measured on M15 and  M20 spectrometers show an increment below T$_c$ for TaRuSi. The curve below T$_c$ is a linear fit to the data, and the constant line above $T_c$ is the average of the points above T$_c$    }
\end{center}
\end{figure*}

 The arc-melted samples of TaReSi and TaRuSi exhibit a normal state mean free path smaller than the coherence length \cite{TXS}, likely placing them in the dirty superconducting limit. The ratio of the superconducting gap values  to $k_BT_c$ in the dirty limit are 1.51(6) and 1.246(30) for TaReSi and TaRuSi, respectively,  are smaller than the weak coupling limit of BCS ratio (namely 1.76) and therefore unphysical and thus requiring explanation going beyond weak-coupling BCS. Possible explanations for this smaller ratio could include the presence of multiple superconducting gaps, a highly anisotropic s-wave gap, or gap nodes (such as would exist for a d-wave gap).  It should be noted that the superconducting state mean free path is unknown and could, in principle, be larger than the normal state value and the superconducting coherence length, which would correspond to clean limit superconductivity.   In the clean limit, TaReSi's superfluid density fits to an isotropic gap ratio of 1.81(5) (near the BCS value), and TaRuSi fits to 1.632(25) (still less than BCS). The measured superfluid density for TaRuSi produces the best goodness of fit for the double gap models with nodes, i.e., s + d or s + p, as shown in Table \ref{fit parameters}. The considerable uncertainty in the magnitude of the d-wave or p-wave gap is a reflection of the fact that the changing relaxation rate $\sigma_{\mathrm{FLL}}$ at low temperature reflects the thermal excitation of normal-state quasiparticles near the gap zeroes, which are not sensitive to the overall gap magnitude. The presence of multiple gaps and the possibility of gap anisotropy which would affect the temperature dependence of $\sigma_{\mathrm{FLL}}$, which makes a definitive statement on the gap symmetry impossible beyond its nodal character:   it is difficult to distinguish whether the gapless state is p or d-wave. Phase-sensitive techniques or measurements sensitive to $\Delta\lambda$ at low temperature ($\mu$SR is sensitive to $1/\lambda^2$) would be better suited for distinguishing between these two possibilities. We note that the low specific heat jump \cite{TXS} observed in TaRuSi is also evidence for a nodal gap.  
 
  \begin{figure}
 \begin{center}
	\includegraphics[width=86mm]{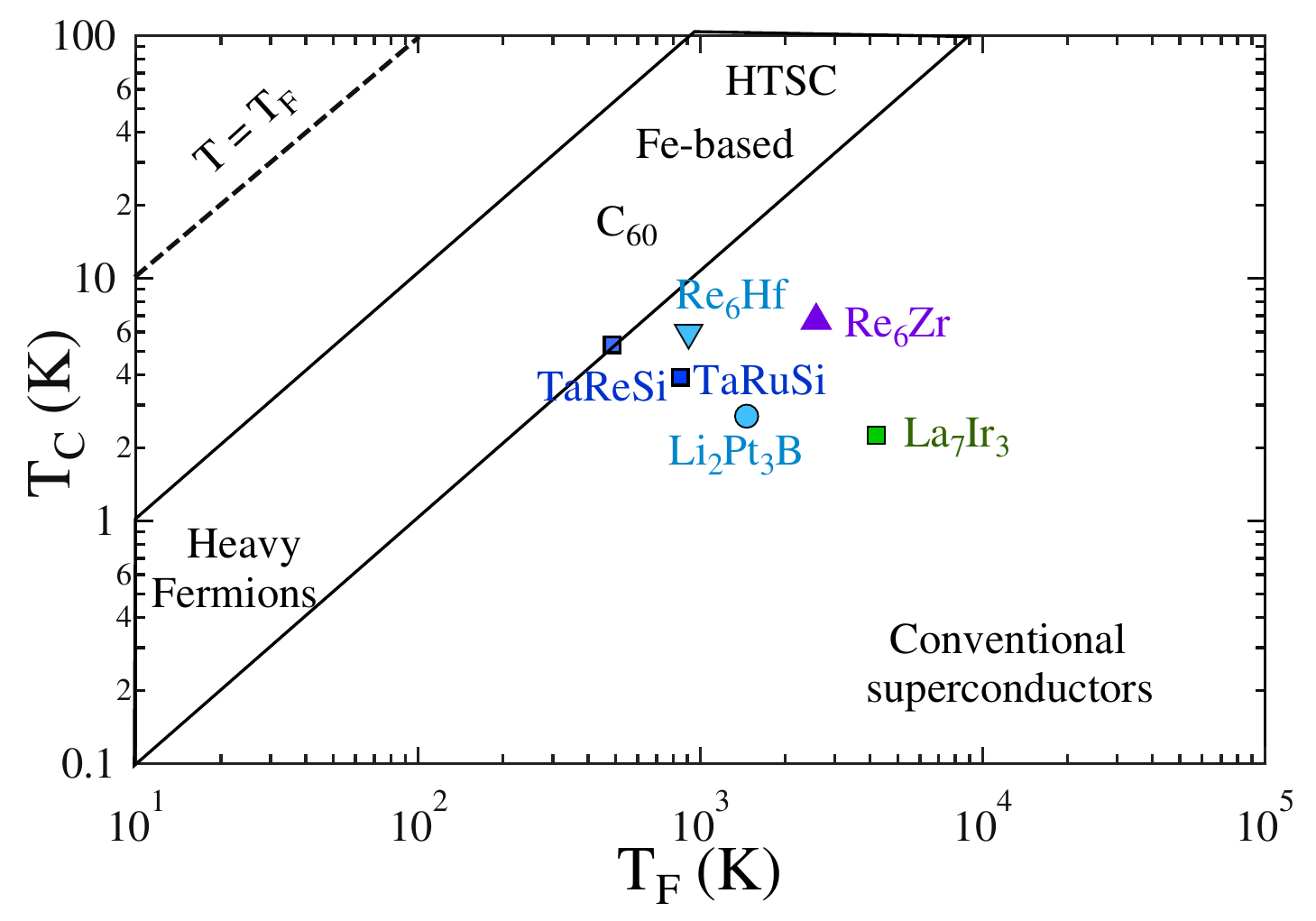}
	\caption{\label{fig3} Log-log plot of $T_c$ versus $T_F$, obtained from relaxation rate ($\sigma_{sc}(0)$) from fig. \ref{fig1}, also called Uemura plot \cite{Ti3X, UemuraPlot, Uemura2}, is shown here. The figure shows that TaRuSi and TaReSi lie close to the band of unconventional superconductors, which include cuprates, iron-based superconductors, heavy fermion superconductors, etc.}
\end{center}
\end{figure}
 Values of the zero temperature penetration depth, $\lambda(0)$, obtained for TaRuSi and TaReSi are 365 nm and 561 nm, respectively, in the clean limit: the values obtained from other models are very similar and can be found in table \ref{fit parameters}. Subsequently, we calculated $n_s(0)/(m^*/m_e)$ which results 2.1$\times$10$^{26}$ m$^{-3}$ and 0.90$\times$10$^{26}$ m$^{-3}$ for TaRuSi and TaReSi, respectively. Here, m$^*$ is the effective mass of the superconducting carrier, and n$_s$ is the superfluid density. Using m$^*$ from ref. \cite{TXS}, the calculated superfluid densities are 0.061 and 0.021 per formulae unit for TaRuSi and TaReSi, respectively. These values are an order of magnitude smaller than the normal state carrier concentrations, which is the case with many unconventional superconductors. The Fermi temperatures, $T_F$, obtained with the help of the Sommerfeld coefficient, $\gamma_n$, from Ref~\cite{TXS} are 848 K and 486 K for TaRuSi and TaReSi, respectively. These T$_F$ values put these superconductors in close proximity to other exotic superconductors on the Uemura plot \cite{UemuraPlot,Uemura2,Ti3X} as shown in Fig.~\ref{fig3}. According to the Uemura classification, the ratio T$_c$/T$_F$ is in the range 0.01 $\leq$ T$_c$/T$_F$ $\leq$ 0.1 for unconventional superconductors and  T$_c$/T$_F$ $\leq$ 0.001 for conventional superconductors.
 
  \begin{table*}
   	\caption{Muon depolarization fitting parameters and resultant superconducting state parameters \newline(c.l. = clean limit, d.l.= dirty limit)}
   	\label{fit parameters}
   	\begin{center}
   		\begin{tabular*}{2.0\columnwidth}{l@{\extracolsep{\fill}}lllllllll}\hline\hline
   			&Model& $\Delta_{0}$(meV) &$T_{c}$(K)&$\Delta_{0}/k_{B}T_{c}$ & fraction& $\chi^{2}$&$\lambda(0)(nm)$&$n_s/(m^*/m_e)$($10^{26} m^{-3}$)\\
   			\hline
   			\\[0.5ex] 
   			\multirow{2}{4em}{TaReSi}& s-wave c.l.& 0.765(15)& 4.89(9)&1.82(5)& & 0.75&561&0.90  \\
   			& s-wave d.l.& 0.627(23)& 4.82(8)&1.51(6)& & 0.78&561&0.90 \\\\
     		\multirow{6}{4em}{TaRuSi}& s-wave c.l. & 0.489(6)&3.47(3)& 1.632(25) & & 0.944&365&2.1  \\
   			& s-wave d.l.& 0.368(8)&3.43(2)& 1.246(30)& & 0.98&364 &2.1 \\
   			& s + s wave& 0.53(7),0.39(20) & 3.46(4)& 1.77(23),1.3(7)& 0.77(66)& 1.02&365&2.1 \\
   			& s + d wave& 0.51(11), 0.5(9)&3.45(3)& 1.7(4), 2(7)& 0.84(65)& 0.79&357&2.2\\
   			& s + p wave& 0.51(1), 0.6(9)&3.45(3)& 1.7(2), 2(3)& 0.84(30)& 0.79&357&2.2\\      
   			
   			\\[0.5ex]
   			\hline\hline
   		\end{tabular*}
   		\par\medskip\footnotesize
   	\end{center}
   \end{table*}
 
 % \begin{table}[!ht]
 %   \centering
 %   \begin{tabular}{llllll}
 %   \hline
%%         & ~ & TaReSi & ~ & TaRuSi & ~ \\ \hline
%        ~ & sc & sd & sc & sd & s+s & s+d \\ 
%        Tc & 4.89(9) & 4.82(8) & 3.69(3) & 3.67(2) & 3.66(3) & 3.66(3) \\ %
%        $\Delta_{0}$ & 0.765(15) & 0.627(23) & 0.592(7) & 0.480(9) & %0.624(25) & 0.614(37) \\ 
%        ~ & ~ & ~ & ~ & ~ & 0.237(128) & 0.793(458) \\ 
%        $\Delta_{0}/k_{B}T_{C}$ & 1.82(5) & 1.51(6) & 1.856(27) & %1.515(30) & 1.98(8) & 1.9(1) \\ 
%        ~ & ~ & ~ & ~ & ~ & 0.8(4) & 3(1) \\ 
%        Fraction & ~ & ~ & ~ & ~ & 0.94 & 0.84 \\ 
%        Chi2 & 0.75 & 0.78 & 1.62 & 1.19 & 1.21 & 1.12 \\ \hline
%    \end{tabular}
%\end{table}
Muon spin relaxation measurements performed in the zero field (ZF) configuration can give unambiguous evidence of a spontaneous magnetic field originating from a TRS breaking superconducting state. ZF spectra for both samples were collected at temperatures above and below the superconducting transition, as shown in Fig. \ref{fig2}(a-b). In the absence of static electronic moments, the muon ensemble polarization decays due to randomly oriented nuclear magnetic moments and is generally described by the Gaussian Kubo-Toyabe function G$_{KT}$(t)

 \begin{equation}
 G_{\mathrm{KT}}(t) = \frac{1}{3}+\frac{2}{3}(1-\sigma^{2}t^{2})\mathrm{exp}\left(-\frac{\sigma^{2}t^{2}}{2}\right) 
 \label{eqn17:zf}
 \end{equation} 
 where $ \sigma $ reflects the width of the nuclear dipolar field experienced by the muons. 

We fit our ZF spectra with the following relaxation function
\begin{equation}
 A(t) = A_{1}G_{\mathrm{KT}}(t)\mathrm{exp}(-\Lambda t)+A_{\mathrm{BG}} 
 \label{eqn18:tay}
 \end{equation}
 where $  A_{1} $ is the sample asymmetry, $  A_{BG} $ is the background asymmetry, and the additional relaxation term exp(-$ \Lambda $t) accounts for any additional relaxation channels (such as broken TRS). The spectra corresponding to TaReSi are seen to relax more rapidly than those for TaRuSi, which can be attributed to the larger nuclear magnetic moment of Re in comparison to Ru.
 
 \begin{figure*}
 \begin{center}
	\includegraphics[width=\textwidth]{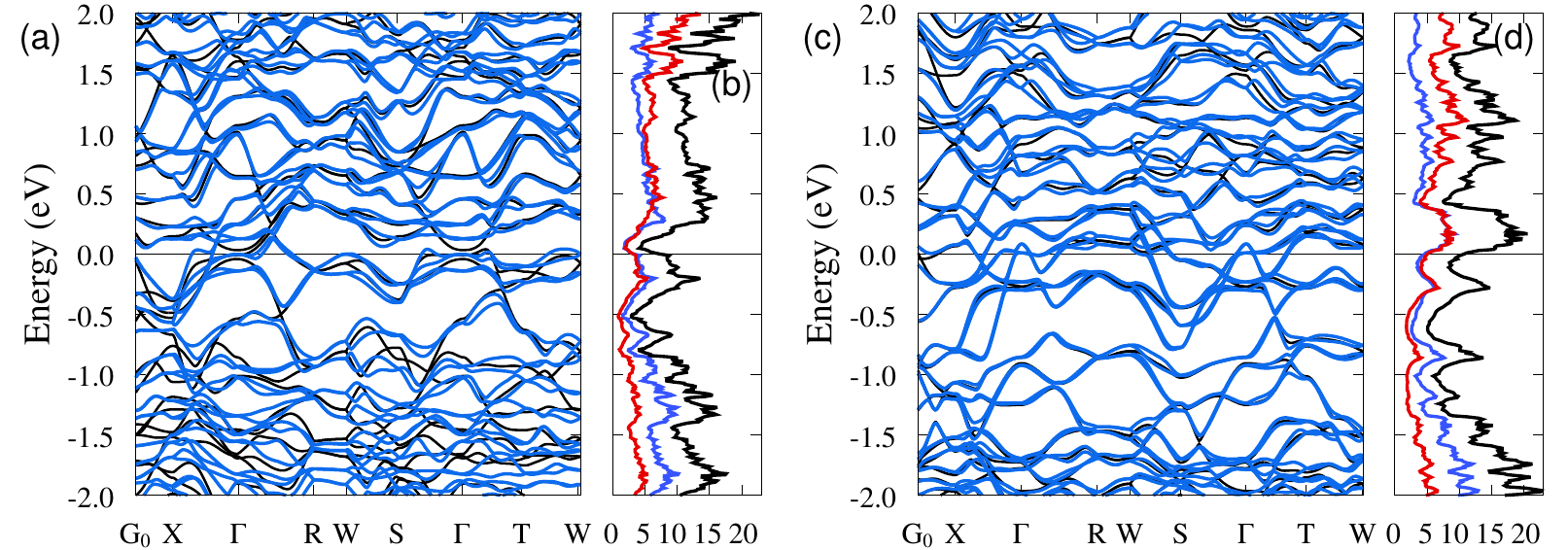}
	\caption{\label{fig4}(a) TaReSi band structure with SOC (blue) and without SOC (black). (b) TaReSi total (black), Re-projected (blue), and Ta-projected (red) density of states with SOC. (c-d) The same quantities were calculated for TaRuSi as in (a-b), respectively.}
\end{center}
\end{figure*}
 In the absence of broken time-reversal symmetry, the ZF-$\mu$SR asymmetry spectra will be temperature independent. However, when there is a spontaneous magnetic field due to  broken time-reversal symmetry in the superconducting state, an additional relaxation of the muon polarization appears below T$_{c}$. In the case of TaRuSi, such a small yet clearly visible difference in the asymmetry spectra can be observed in Fig. \ref{fig2}(b),(e) while the spectra for TaReSi are temperature independent  as evident from Fig. \ref{fig2}(a),(d). First, we performed the analysis of asymmetry spectra using equation \ref{eqn18:tay} where $\Lambda$ was kept temperature-independent while the  Kubo-Toyabe relaxation rate $\sigma$ was allowed to vary with temperature in accordance with Ref. \cite{SRO, UP}; these results are plotted in Fig. \ref{fig2}(a-c). Secondly, we analysed the asymmetry spectra temperature-dependent $\Lambda$ and temperature-independent $\sigma$  as depicted in Fig. \ref{fig2}(d-e).  Both models describe the data well for TaReSi. For, TaReSi, the fitted values of $\Lambda$  and $\sigma$ are temperature-independent as shown in the inset of Fig.~\ref{fig2}(a),(d) characteristic of  preserved time-reversal symmetry. We estimated the maximum value of the magnetic field that could exist in TaReSi within our fit uncertainties to be 0.003 mT. 
  
  %However, for TaRuSi the fitting with temperature-independent $\sigma$ and temperature-dependent $\Lambda$ doesn't describe the experimental data well at longer times as shown in Fig. \ref{fig2}(e), also, evident from goodness of fit parameter ($\chi^2$=1.027). We found that using a temperature-independent exponential relaxation rate $\Lambda$ and temperature-dependent Kubo-Toyabe relaxation rate $\sigma$  produces statistically better fits with better goodness of fit  ($\chi^2$=1.024)  .  
  We note that $\sigma$ and $\Lambda$ are highly correlated when the relaxation is weak  and can therefore trade off against each other when allowed to vary simultaneously, obscuring the observation of any increase in overall relaxation. We observe a small but clear increase in the Kubo-Toyabe relaxation rate $\sigma$ for TaRuSi as the temperature is decreased below T$_{c}$ as shown in Fig.~\ref{fig2}{c}. Similarly, we observed a increase in relaxation rate parameter $\Lambda$  when $\sigma$ was kept temperature independent as shown in Fig.~\ref{fig2}(f) . A temperature-independent value of the relaxation rates above T$_{c}$  with an increase below T$_{c}$ , irrespective of the fitting method, indicates the onset of a small spontaneous magnetic field in the superconducting state for TaRuSi, a behavior characteristic of broken time-reversal symmetry in the superconducting state.  The increase in relaxation is consistent with an onset temperature of the superconducting T$_c$, however, we cannot make a definitive statement on the precise onset temperature due to the relatively large error bars in the relaxation rate. We note that neither our transverse field measurements or any bulk measurements (susceptibility, resistivity, specific heat) show evidence for multiple phase transitions, making it most likely that the relaxation increases from T$_C$.  From the average increase in the ZF relaxation rate,%($\Delta \sigma$ $\approx$ 0.062 $\mu$s$^{-1}$),%
 the magnitude of the spontaneous field can be estimated as $\sqrt{2}\Delta \sigma$/$\gamma_{\mu} \approx$ 0.010~mT. %The maximum value of the field that can exist in TaRuSi is 0.012~mT.%
 This value is less than that reported for TRS-breaking superconductors UPt$_{3}$ \cite{UP} and Sr$_{2}$RuO$_{4}$ \cite{SRO}; however, it is comparable to weakly correlated non-centrosymmetric superconductors Re$_{6}$X (X = Zr, Hf, Ti) \cite{RZ3,RH2,RT}, La$_{7}$X$_{3}$ (X = Ir, Rh, Ni) \cite{LI,LaRh,LN} and LaNiC$_2$ \cite{LNC1}. Related works published during the review process of this manuscript  also report time-reversal symmetry breaking superconducting state in TaRuSi \cite{tshangTaRuSi} and apparently time reversal symmetry preserving superconducting state in TaReSi \cite{tshangTaReSi}, in agreement with our results presented here.
 
  \begin{figure*}
 \begin{center}
	\includegraphics[width=\textwidth]{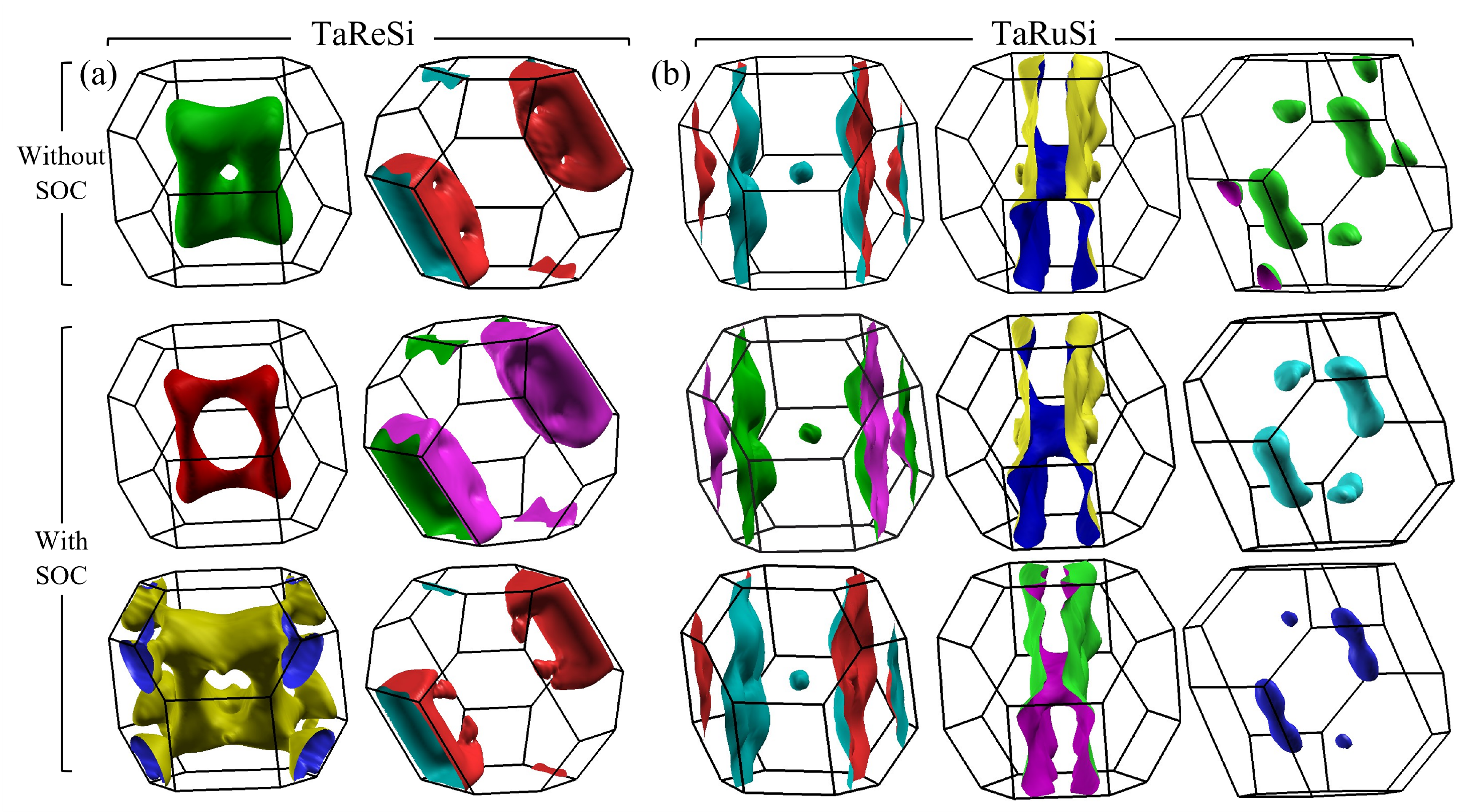}
	\caption{\label{fig5} Fermi surface plots for TaReSi and TaRuSi with and without SOC. The effect of SOC on splitting the Fermi surfaces can be seen by comparing the top row to the bottom two rows in the case of (a) TaReSi and (b) TaRuSi. Fermi surface sections plotted in (a) differ significantly between the top and bottom rows, indicating the strong effect of SOC on Fermi surface splitting for TaReSi. However, the Fermi surface sections plotted in the top row of (b) are close to those of the bottom rows, indicating a weak effect of SOC on Fermi surface splitting for TaRuSi.}
\end{center}
\end{figure*}
 
To further characterize TaXSi, we have performed fully-relativistic electronic structure calculations using the plane-wave pseudopotential formalism of density functional theory (DFT) as implemented in the Quantum Espresso code \cite{QE}. For the exchange-correlation functional, we have used the generalized gradient approximation \cite{PBE_1996}. A 6$\times$6$\times$6 Monkhorst-Pack $\vec{k}$-point grid was used to sample the Brillouin zone in the DFT self-consistency cycle, while the density of states and Fermi energy were computed using a dense 24$\times$24$\times$24 $\vec{k}$-point grid. We have used a kinetic energy cutoff of 100 Ry for the wavefunctions and 400 Ry for the charge density. Projector augmented-wave pseudopotentials with nonlinear core correction, as provided in the PSlibrary, were used. The unit cell and atomic positions are relaxed until forces on each atom are less than 10$^{-3}$ Ry/Bohr. The experimentally determined lattice constants \cite{TXS} were used as a starting condition for the relaxation. The resulting lattice constants are $a=7.02$~\AA, $b=11.56$~\AA, and $c=6.70$~\AA\ for TaReSi, and $a=7.30$~\AA, $b=11.20$~\AA, and $c=6.51$~\AA~ for TaRuSi are in excellent agreement with the values reported in Ref. \cite{TXS} which were measured using X-ray diffraction at room temperature.

Fig.~\ref{fig4} shows the computed DFT band structure for both materials with and without SOC as well as the total and atom-projected density of states. Fermi surface sections obtained from our DFT calculations with and without SOC are shown in Fig.~\ref{fig5}. For both compounds, we have found multiple Fermi surface sheets, allowing for the possibility of multi-gapped superconductivity. Fig.~\ref{fig4} also shows that bands near the Fermi energy for TaReSi are significantly altered by SOC, while bands near the Fermi energy for TaRuSi are only slightly altered by the inclusion of SOC. Similarly, by comparing the top and bottom rows of plots in Fig.~\ref{fig5}, it can be seen that there is a large splitting of Fermi surface sections due to SOC for TaReSi, and a relatively much weaker splitting of Fermi surface sections for TaRuSi.

A common measure of the ASOC energy, $E_{\rm ASOC}$, is obtained through estimating the band splitting near the Fermi energy \cite{Smidman_Agterberg_2017}. It can be seen in Fig.~\ref{fig4} that, for TaRuSi, the band splitting is at most about 50 meV, while for TaReSi, the band splitting is well over 100 meV for multiple bands near the Fermi energy. We have also computed the band splitting for each band which crosses the Fermi energy over $32\times32\times32$ $\vec{k}$-points in the first Brillouin zone. For TaRuSi, we find that the band splitting is at most 100-150 meV only for $\vec{k}$-points on a few small pockets of the Fermi surface, and less than 70 meV for all $\vec{k}$-points comprising the larger branches of the Fermi surface. We have not observed a large band splitting of 300 meV as stated in Ref.\cite{tshangTaRuSi} at any $\vec{k}$-point within the first Brilluoin zone for any of the bands which cross the Fermi energy. In contrast, for TaReSi, much of the band splitting for $\vec{k}$-points across the entire Fermi surface is greater than 100 meV. Using the band splittings which we have computed across the first Brillouin zone, we have computed the average band splitting for each system, which we refer to as $E_{\rm ASOC}$. In computing $E_{\rm ASOC}$, we only consider $\vec{k}$-points at which one of the split bands is above the Fermi energy and the other below the Fermi Energy. We have found that $E_{\rm ASOC} = 41$~meV for TaRuSi and $E_{\rm ASOC} = 81$~meV for TaReSi. Our results show that the effects of SOC on the band structure for TaRuSi (TaReSi) are relatively weak (strong) when compared with other NSC superconductors. For example, the band splittings for TaRuSi and TaReSi are comparable to those of LaNiC$_2$~\cite{Hase_Yanagisawa_2009} and CePt$_3$Si~\cite{Samokhin_Bose_2004} respectively.

%We have computed the average band splitting for each system over $32\times32\times32$ $\vec{k}$-points in the first Brillouin zone which we refer to as $E_{\rm ASOC}$. In computing $E_{\rm ASOC}$, we only consider $\vec{k}$-points at which one of the split bands is above the Fermi energy and the other below the Fermi Energy. We have found that $E_{\rm ASOC} = 41$~meV for TaRuSi and $E_{\rm ASOC} = 81$~meV for TaReSi. We have also checked the band splitting for each $\vec{k}$-point across the Fermi surfaces. For TaRuSi, we find that the band splitting is at most 100-150 meV only for a few small pockets of the Fermi surface, and less than 50 meV for the larger branches of the Fermi surface. We have not observed a large band splitting of 300 meV as stated in Ref.\cite{tshangTaRuSi} at any $\vec{k}$-point within the first Brilluoin zone for any of the bands which cross the Fermi energy. In contrast, for TaReSi, much of the band splitting across the entire Fermi surface is greater than 100 meV. The calculated band splittings for TaRuSi and TaReSi are comparable to LaNiC$_2$ and CePt$_3$Si~\cite{Samokhin_Bose_2004} respectively.

The electron-phonon coupling strength is estimated through the renormalization factor, $1+\lambda_{\rm e-ph}$ \cite{Marsiglio_Carbotte_2008}, required to match the computed density of states at the Fermi energy, $D(E_{\rm F})$, to the values of 2.28 and 3.34 $\frac{\rm states}{\rm eV\ f.u.}$ obtained through measurements of the Sommerfeld constant for TaReSi and TaRuSi respectively \cite{TXS}. From the calculated density of states, we have obtained $D(E_{\rm F}) = 1.26$ and $1.78$ $\frac{\rm states}{\rm eV\ f.u.}$ giving the estimates $\lambda_{\rm e-ph} = 0.81$ and $0.88$ for TaReSi and TaRuSi respectively. Both values are close to the values obtained previously from the McMillan formula \cite{TXS} and indicate moderate electron-phonon coupling in both materials. From the atom-projected DOS, we observe that Ta and Re/Ru orbitals are hybridized, with each contributing equally to conduction.

The results we have obtained for TaRuSi  can be compared with what has been observed for LaNiC$_2$. Both systems break TRS and share the same point group, $C_{2v}$. Therefore, the symmetry analysis of Ref.~\cite{Quintanilla_Cywinski_2010} is applicable to TaRuSi as well. In particular, since $C_{2v}$ has only one-dimensional irreducible representations, TRS {\it breaking} in the superconducting state is only possible when SOC is weak. This is compatible with the small band splitting due to SOC which we have obtained from our DFT calculations for TaRuSi. It is also consistent with the {\it preservation} of TRS observed in TaReSi, for which we have shown that there is significant band splitting due to SOC. Furthermore, like TaReSi, ThCoC$_2$ has the same point group, $C_{2v}$, DFT calculations indicate large band splitting due to SOC of 150 meV averaged over the first Brillouin zone \cite{Kuderowicz_Wojcik_2021}, and TRS is preserved in the superconducting state \cite{Bhattacharyya_Manfrinetti_2019}. Following the analysis of Ref.~\cite{Quintanilla_Cywinski_2010}, only a non-unitary triplet-pairing state is compatible with our observations of TRS breaking in the superconducting state of TaRuSi. All such states necessarily exhibit gap nodes, consistent with our penetration depth measurements. It is suspected that Hund's coupling between electrons on Ni orbitals could provide the pairing mechanism for non-unitary triplet pairing in LaNiC$_2$ \cite{Csire_Annett_2018} and LaNiGa$_2$ \cite{Weng_Yuan_2016, Ghosh_Quintanilla_2020}. Similar arguments could be made for describing the possible triplet-pairing mechanism in TaRuSi given Hund's coupling is also significant in many Ru-based materials \cite{Dang_Millis_2015, Georges_Mravlje_2013}.

%There are a few notable differences between the intrinsic properties of TaReSi and TaRuSi which could provide a better understanding of the distinct superconducting states. The effects of ASOC on the electronic structure are clearly much stronger in the Re-based system. 
It was speculated that large ASOC in Li$_2$Pt$_3$B relative to that of Li$_2$Pd$_3$B could explain significant spin-triplet pairing and line nodes in the energy gap of Li$_2$Pt$_3$B, while Li$_2$Pd$_3$B, which has smaller ASOC, exhibits spin-singlet pairing and an s-wave gap \cite{LPB2}. Similarly to the case of Li$_2$Pd$_3$B, electronic structure calculations for BaPtSi$_3$ have shown a small band splitting near the Fermi surface due to ASOC \cite{Bauer_Marsman_2009} while specific heat measurements are in line with what is expected for a fully-gapped spin-singlet-pairing superconductor \cite{Bauer_Marsman_2009}. This, along with our results, suggests that strong ASOC can drive superconducting systems to unconventional or conventional superconductivity for some underlying crystal structure symmetries.   
%It is possible that electronic correlations may play a role in explaining the observed differences between TaReSi and TaRuSi.

In conclusion, we have studied two iso-structural superconductors, TaRuSi and TaReSi, via zero-field and transverse-field $\mu$SR and DFT calculations. We have shown that the inclusion of ASOC in our DFT calculations significantly modifies the electronic structure of TaReSi when compared with DFT calculations neglecting ASOC, producing a large band splitting of over 100 meV for multiple bands near the Fermi energy. In contrast, the electronic structure of TaRuSi is only weakly affected by the inclusion of ASOC, with an average band splitting of about 40 meV. Given that each of these compounds share the point-group symmetry, $C_{2v}$, and that TaRuSi is shown to break TRS in the superconducting state while TaReSi does not, our results are in alignment with the symmetry analysis of Ref.~\cite{Quintanilla_Cywinski_2010} for superconductors with the $C_{2v}$ point-group symmetry. That is, we have only observed TRS breaking in the case of weak SOC, leading us to conclude that  the superconducting state for TaRuSi is a non-unitary triplet-pairing state. TaRuSi would therefore be the first system other than LaNiC$_2$ with the point group $C_{2v}$ to host such a non-unitary triplet-pairing state. Our fits of the transverse field $\mu$SR data indicate fully-gapped superconductivity in TaReSi while we find  superconductivity with gap nodes in TaRuSi, which is expected from symmetry analysis \cite{Quintanilla_Cywinski_2010} for a broken time reversal symmetry state.

Work at McMaster was supported by the Natural Sciences and Engineering Research of Council of Canada. R. P. S. acknowledges the Science and Engineering Research Board, Government of India, for the Core Research Grant CRG/2019/001028. The financial support from DSTFIST Project No. SR/FST/PSI-195/2014(C) is also thankfully acknowledged. This research was enabled in part by support provided by SHARCNET (sharcnet.ca) and the Digital Research Alliance of Canada (alliancecan.ca).

\bibliography{TaXSi}

\end{document}